# Strain and Water Effects on the Electronic Structure and Chemical Activity of In-Plane Graphene/Silicene Heterostructure

Andrey A Kistanov[1,2,*], Yongqing Cai[2], Yong-Wei Zhang[2,*], Sergey V Dmitriev[3,4] and Kun Zhou[1,*]

[1]School of Mechanical and Aerospace Engineering, Nanyang Technological University, 639798 Singapore

[2]Institute of High Performance Computing, Agency for Science, Technology and Research, 138632 Singapore

[3]Institute for Metals Superplasticity Problems, Russian Academy of Sciences, 450001 Ufa, Russia

[4]Research Laboratory for Mechanics of New Nanomaterials, Peter the Great St. Petersburg Polytechnical University, 195251 St. Petersburg, Russia

**E-mail**: kzhou@ntu.edu.sg (K. Z.), andrei.kistanov.ufa@gmail.com (A. A. K.) and zhangyw@ihpc.a-star.edu.sg (Y.-W. Z.)

## Abstract

By using first-principles calculations, the electronic structure of planar and strained in-plane graphene/silicene heterostructure is studied. The heterostructure is found to be metallic in a strain range from -7% (compression) to +7% (tension). The effect of compressive/tensile strain on the chemical activity of the in-plane graphene/silicene heterostructure is examined by studying its interaction with the $H_2O$ molecule. It shows that compressive/tensile strain is able to increase the binding energy of $H_2O$ compared with the adsorption on a planar surface, and the charge transfer between the water molecule and the graphene/silicene sheet can be modulated by strain. Moreover, the presence of the boron-nitride (BN)-substrate significantly influences the chemical activity of the graphene/silicene heterostructure upon its interaction with the $H_2O$ molecule and may cause an increase/decrease of the charge transfer between the $H_2O$ molecule and the heterostructure. These findings provide insights into the modulation of electronic properties of the in-plane free-standing/substrate-supported graphene/silicene heterostructure, and render possible ways to control its electronic structure, carrier density and redox characteristics, which may be useful for its potential applications in nanoelectronics and gas sensors.

**Keywords:** graphene, silicene, in-plane heterostructure, chemical activity, strain and water effects, BN-substrate

# 1. Introduction

Since its first fabrication in 2004 [1], graphene, a two-dimensional (2D) single atomic layer of crystalline carbon, has been intensively investigated because of its unique opto-electronic and mechanical properties, such as a high carrier mobility [2] of 200,000 cm$^2$/Vs, an extraordinarily high Young's modulus [3] of 1.0 TPa and extra flexibility [4]. Recently produced silicene [5], a 2D allotrope of silicon, is also an important material among other 2D materials. Similar to graphene, silicene is a zero-gap semiconductor with a tunable band gap [6]. In addition, silicene possesses a high intrinsic carrier mobility [7] of 2.57·10$^5$ cm$^2$/V·s, and suitable mechanical flexibility [8]. These extraordinary properties suggest silicene as another promising material for novel device applications. However, constraints like the absence of the band gap, and poor mechanical properties of silicene have driven researchers to seek other novel materials. Over the past few years, many researches have been focused on hybrid 2D materials, such as graphene/hexagonal-boron-nitride (h-BN) [9, 10], graphene/transition metal dichalcogenide (TMD) [11, 12], h-BN/silicene [13], and graphene/silicene heterostructures [14]. These 2D heterostructures are designed to overcome limitations and to develop the performance of individual 2D materials [15−19].

Furthermore, because of its atomically thin structure and large surface area, 2D materials such as silicene are easily subjected to exposure of the environment, and their electronic properties can be greatly affected by it [20, 21]. To protect chemically unstable 2D materials, commonly, passivation by using more stable 2D materials, such as graphene or h-BN, as a capping layer is used [22, 23]. However, manufacturing of vertically stacked materials has disadvantages, due to the possible contamination between layers, which leads to significant challenges for massive production of the samples [24, 25]. In that case, the in-plane interconnected heterostructures, for which there are no such issues in their production, have attracted great attentions from both theoretical and experimental sides [26, 27].

Recently, many studies have reported on the effect of various factors on the electronic properties of different heterostructures [9, 28−30]. For example, electric-field engineering was applied to modify the band gap of graphene/h-BN heterostructures [31]. The works [32, 33] predicted that despite that the differences in the electronic and mechanical properties of graphene/silicene heterostructure under different types of applied strain may exist, for a gas adsorption at sites far away from the interface region, the type of strain applied is expected to exert little effect due to the nearly isotropic nature of both graphene and silicene sheets. Doping of graphene/silicon junction solar cells led to a significant enhancement, up to 3.9%, of energy conversion efficiency [34]. Strain engineering was found as an effective way to change the band gap of heterostructures. In addition, mechanically-tuned band gap was predicted in graphene/h-BN [30, 31, 34, 35] and graphene/$MoS_2$ heterostructures [36, 37]. However, it is noted that the strain effects on the in-plane graphene/silicene heterostructure remain unexplored.

In this work, we perform investigations on the effects of strain and adsorption of humidity ($H_2O$) molecule on the electronic structure and chemical activity of the in-plane graphene/silicene heterostructure. The considered heterostructure is found to be metallic with the Dirac point above the Fermi level, which can be shifted by applying either compressive or tensile strains. Moreover, the chemical activity of graphene/silicene heterostructure under the compressed (-7%) and tensile (7%) strains is found to be significantly large in comparison with that of its flat counterpart. The analysis is also conducted on the effect of the BN-substrate on the adsorption energies and the charge transfer between the graphene/silicene heterostructure and the adsorbed $H_2O$ molecule. It shows that the BN-substrate significantly influences the donor/acceptor ability of $H_2O$ molecule upon its adsorption on the graphene/silicene heterostructure and may cause an increase/decrease of the charge transfer between the $H_2O$ molecule and heterostructure. The present study provides new knowledge on the properties of the in-plane graphene/silicene heterostructure and reveals the effects of strain and $H_2O$ molecules on its electronic and chemical properties, which may be useful for its potential device applications.

## 2. Computational methods

All the calculations are based on the density functional theory (DFT) and performed by spin-polarized first-principles calculations. Vienna *ab initio* simulation package (VASP) [38] with the Perdew-Burke-Ernzerhof (PBE) functional [39] under the generalized gradient approximation (GGA) is used. The van der Waals (vdW) corrected functional with Becke88 optimization (optB88) [40] is selected to analyze dispersive interactions during the noncovalent chemical functionalization of structures with small molecules. All the structures are fully relaxed until the total energy and atomic forces were smaller than $10^{-5}$ eV and 0.01 eV/Å, respectively.

The free-standing graphene/silicene heterostructure is created by using the $5 \times 5 \times 1$ and $3 \times 3 \times 1$ supercells of graphene and silicene (60 carbon and 24 silicon atoms), respectively. To create BN-supported graphene/silicene heterostructure the initial free-standing heterostructure is placed on the BN-substrate. The substrate is created by using the $5 \times 11 \times 1$ supercell (60 boron and 60 nitride atoms). Periodic boundary conditions are applied in the two in-plane transverse directions, while the vacuum space of 20 Å is introduced along the out-of-plane direction. Due to the difference between the lattice spacing of graphene ($a_{gr}$) and that of silicene ($a_{si}$), the mismatch strain along the interface of the graphene/silicene heterstructure is $\varepsilon_{mismatch} = (5a_{gr} - 3a_{si})/(5a_{gr}) \approx 2.7\%$. Since during the relaxation, the supercell period along the zigzag direction is taken as $5a_{gr}$, the strain along the interface direction in the graphene is $\varepsilon_{gr} = 0\%$, and that in the silicene is $\varepsilon_{si} = \varepsilon_{mismatch}$. The compressed/stretched structure is obtained by applying uniaxial strain in the direction perpendicular to the interface of the graphene/silicene heterostructure. The choice of the asymmetric graphene/silicene interface is based on previous works [41, 42] of in-plane heterostructures, where epitaxy of graphene with other 2D material occurs preferentially along the zigzag direction.

The first Brillouin zone is sampled with a $10 \times 10 \times 1$ k-mesh grid. The kinetic energy cutoff of 450 eV is adopted. The adsorption energy $E_a$ of a molecule on the free-standing and BN-substrated graphene/silicene heterostructure is calculated as $E_a = E_{Mol+H} - E_{Mol} - E_H$, where $E_{Mol+H}$, $E_{Mol}$ and $E_H$ are the energies of the free-standing/BN-supported molecule adsorbed graphene/silicene heterostructure, molecule, and the free-standing/BN-supported graphene/silicene sheet, respectively. The electronic interaction between the $H_2O$ molecule with graphene/silicene heterostructure is analyzed by differential charge density (DCD) $\Delta\rho(r)$ defined as the difference between the total charge density of the molecular-adsorbed graphene/silicene system and the sum of the charge densities of the isolated molecule and the planar graphene/silicene heterostructure. The exact amount of charge transfer between the molecule and the surface, the plane-averaged DCD $\Delta\rho(z)$ along the normal direction z of the graphene/silicene surface, is calculated by integrating $\Delta\rho(r)$ over the basal plane at the z point. The amount of transferred charge at the z point is given by $\Delta Q(z) = \int_{-\infty}^{z} \Delta\rho(z')dz'$. Based on the $\Delta Q(z)$ curves, the total amount of charge donated by the molecule is read at the interface between the molecule and the graphene/silicene where $\Delta\rho(z)$ shows a zero value. The geometric structures and charge density distributions are plotted through Visualization for Electronic and Structural Analysis (VESTA) [43].

## 3. Results and discussion

### 3.1 Strain effect on the electronic structure

Figures 1(a) − (e) present the variation of band gap of the in-plane graphene/silicene heterostructure under the tension strain of 7% and 5%, planar structure (0% of strain), and under the compressive strain of -5% and -7%, respectively, along the armchair direction. It is seen that the heterostructure remains a metal within the strain

range as there is no band gap near the Fermi level. For each considered case, the Dirac point is located above the Fermi level, signifying a p-type of conduction.

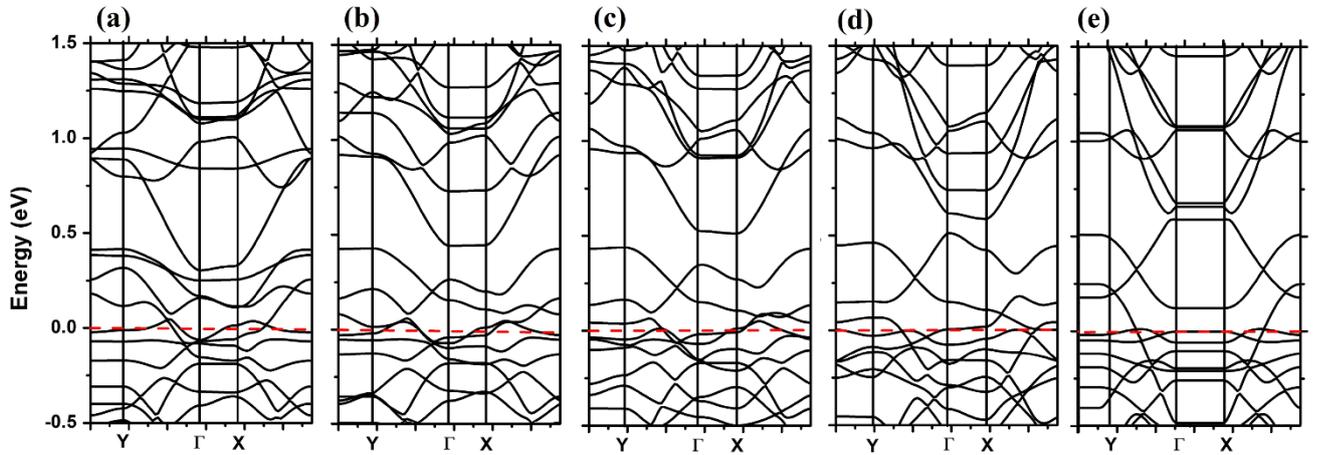

**Figure 1.** Variation of band gap of in-plane graphene/silicene heterostructure under different strains applied along the armchair direction (a)-(e). The applied strains are (a) 7%, (b) 5%, (c) 0%, (d) -5% and (e) -7%, respectively.

### 3.2 Interaction with environmental $H_2O$ molecule

The adsorption of the environmental $H_2O$ molecule on the graphene, silicene, and graphene/silicene interfacial regions of the in-plane graphene/silicene surface is considered. For each case, several possible symmetric anchoring positions of the molecule on the planar and compressed surfaces are examined. All subsequent calculations on the electronic properties and energetics are based on the lowest-energy configurations of the adsorbed heterostructure. The most stable configurations for the case of $H_2O$ molecule adsorbed on the planar sheet are given in figures 2(a), (d) and (j). In case of $H_2O$ adsorbed on the graphene region (figure 2(a)), both of the O−H bonds are disposed at the angle of around 45° to the surface and located directly above the ridge of graphene. The distance from the molecule to the surface is 2.87 Å, and the value of $E_a$ is −0.152 eV, which is consistent with the result reported in [44]. Figure 2(d) shows the $H_2O$ molecule adsorbed on silicene region, in which one of the O−H bonds is parallel to the surface along the armchair direction and the other nearly normal to the surface. The distance from the molecule to the surface is 2.89 Å, and $E_a$ is −0.140 eV, which is in consistent with result in [45]. Figure 2(j) presents the $H_2O$ adsorbed on the graphene/silicene region. It is seen that the molecule is located directly above the ridge of the graphene/silicene site, both O−H bonds are nearly parallel to the surface, the distance between the molecule and the surface is 2.71 Å, and $E_a$ is −0.175 eV.

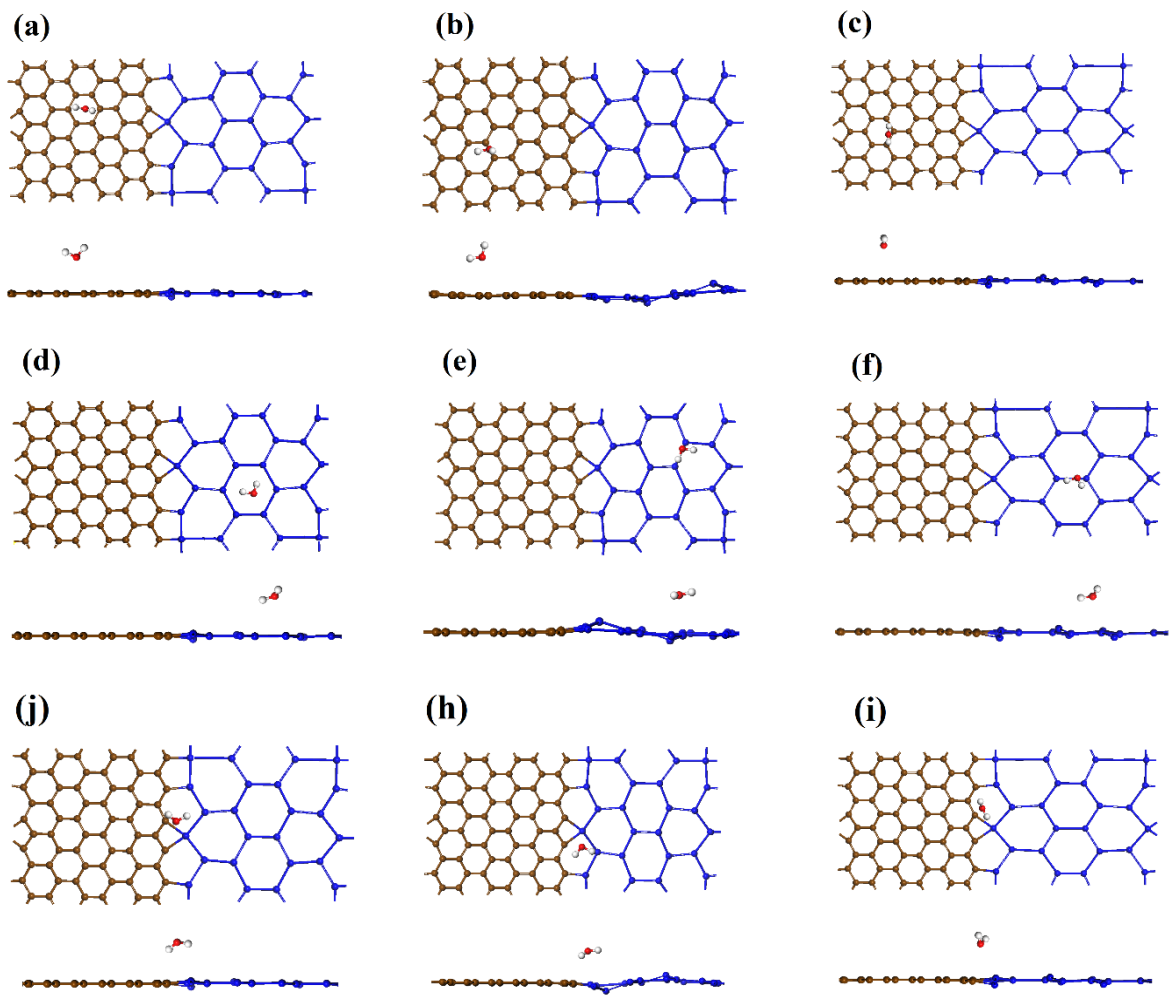

**Figure 2.** The most stable adsorption positions of the H$_2$O molecule on different regions of the heterostructure. Under 0% strain: (a) graphene region, (d) silicene region, and (j) the graphene/silicene interfacial region. Under -7% strain: (b) graphene region, (e) silicene region, and (h) the graphene/silicene interfacial region. Under 7% strain: (c) graphene region, (f) silicene region, and (i) the graphene/silicene interfacial region, respectively. The balls in brown, blue, white and red represent carbon, silicon, hydrogen and oxygen atoms, respectively.

For the H$_2$O adsorbed on the graphene/silicene sheet under the compressed strain of -7%, the most stable configurations are given in figures 2(b), (e) and (h). Figure 2(b) shows the H$_2$O adsorbed on the graphene region, in which the in-plane O−H bond is parallel to the surface along the armchair direction while the out-of-plane O−H bond is nearly normal to the surface and located directly above the armchair C−C bond. The distance from the molecule to the surface is 2.97 Å, and the value of $E_a$ is −0.157 eV. In the case of the H$_2$O adsorbed on the silicene region (figure 2(e)), both O−H bonds are parallel to the surface and located directly above the zigzag Si−Si bond. The distance from the molecule to the surface is 2.82 Å, and $E_a$ is −0.175 eV. Figure 2(h) presents the H$_2$O adsorbed on the graphene/silicene interfacial region, in which the molecule is located directly above the ridge of the graphene/silicene site and both of the O−H bonds are nearly parallel to surface, the distance between the molecule and the surface is 2.40 Å, and $E_a$ is −0.210 eV.

For the $H_2O$ adsorbed on the graphene/silicene sheet under the tensile strain of 7%, the most stable configurations are given in figures 2(c), (f) and (i). Figure 2(c) shows the $H_2O$ adsorbed on the graphene region, in which the O−H bonds are disposed at the angle of around 45° to the surface and located directly above the armchair C−C bonds. The distance from the molecule to the surface is 3.03 Å, and the value of $E_a$ is −0.145 eV. In the case of the $H_2O$ adsorbed on the silicene region (figure 2(f)), both O−H bonds are parallel to the surface and located directly above the armchair Si−Si bond. The distance from the molecule to the surface is 2.72 Å, and $E_a$ is −0.154 eV. Figure 2(i) presents the $H_2O$ adsorbed on the graphene/silicene interfacial region, in which the molecule is located directly above the ridge of the graphene/silicene site and both of the O−H bonds are nearly parallel to surface, the distance between the molecule and the surface is 2.61 Å, and Ea is −0.160 eV.

It is found that in the case of the compressive strain the adsorption energy $E_a$ increases with strain. A large enhancement is observed for the cases of $H_2O$ adsorption on the silicene and graphene/silicene interfacial regions, which may be explained by the large distortion of the silicene lattice and Si-Si bonds deformation [46]. Tensile strain leads to the decrease of the adsorption energy $E_a$ in the cases of the $H_2O$ molecule adsorption on the graphene and interfacial graphene/silicene regions, while increases in the case when $H_2O$ adsorbed on the silicene region. Table 1 summarizes the energetics data of the $H_2O$ molecular adsorption.

**Table 1.** Adsorption energy ($E_a$) and charge transfer ($\Delta q$) from $H_2O$ molecules to different regions of the planar, compressed and stretched graphene/silicene heterostructure, as well as the X−H bond length ($B_{X-H}$), where X represents the $H_2O$ molecule. Note that a positive $\Delta q$ indicates the transfer of electrons from the molecules to the graphene/silicene heterostructure.

| Strain | Heterostructure regions | $E_a$ (eV) | $\Delta q$ (e) | $B_{x-h}$ (Å) | Molecule |
|---|---|---|---|---|---|
| 0% | graphene | -0.152 | 0.024 | 2.87 | donor |
| | silicene | -0.140 | -0.092 | 2.89 | acceptor |
| | graphene/silicene | -0.175 | -0.057 | 2.71 | acceptor |
| -7% | graphene | -0.157 | 0.035 | 2.97 | donor |
| | silicene | -0.175 | 0.181 | 2.82 | donor |
| | graphene/silicene | -0.210 | -0.355 | 2.67 | acceptor |
| 7% | graphene | -0.145 | -0.028 | 3.03 | acceptor |
| | silicene | -0.154 | 0.160 | 2.72 | donor |
| | graphene/silicene | -0.160 | -0.050 | 2.61 | acceptor |

To gain insight into the electronic properties of the planar (under 0% strain), compressed (under the strain of -7%) and stretched (under the strain of 7%) graphene/silicene heterostructure after $H_2O$ doping, we study the local density of states (LDOS). The LDOS analysis reveals that the additional electronic states induced by $H_2O$ are located below the Fermi level for the molecular adsorption on planar, compressed or stretched surfaces (figure 3). However, the alignment of the molecular levels of $H_2O$ is strongly dependent on the region of molecular adsorption and also affected by strain.

Figures 3(a), (d) and (j) present the three highest occupied molecular orbitals $1b_1$ (HOMO), $3a_1$ (HOMO-1), and $1b_2$ (HOMO-2) of the $H_2O$ molecule adsorbed, respectively, on the graphene, silicene and graphene/silicene interfacial regions of the planar sheet. Clearly, the distributions of molecular orbitals are different for each region of the heterostructure. Moreover, by comparing the planar, compressed and stretched structures of silicene (as shown in figures 3(b), (e) and (h), respectively) and the graphene/silicene interfacial region (figures 3(c), (f) and (e)), we see that $1b_1$, $3a_1$, and $1b_2$ molecular orbitals of the $H_2O$ molecule are shifted downwardly by around 0.25 eV for the cases when the compressive or tensile strains are applied.

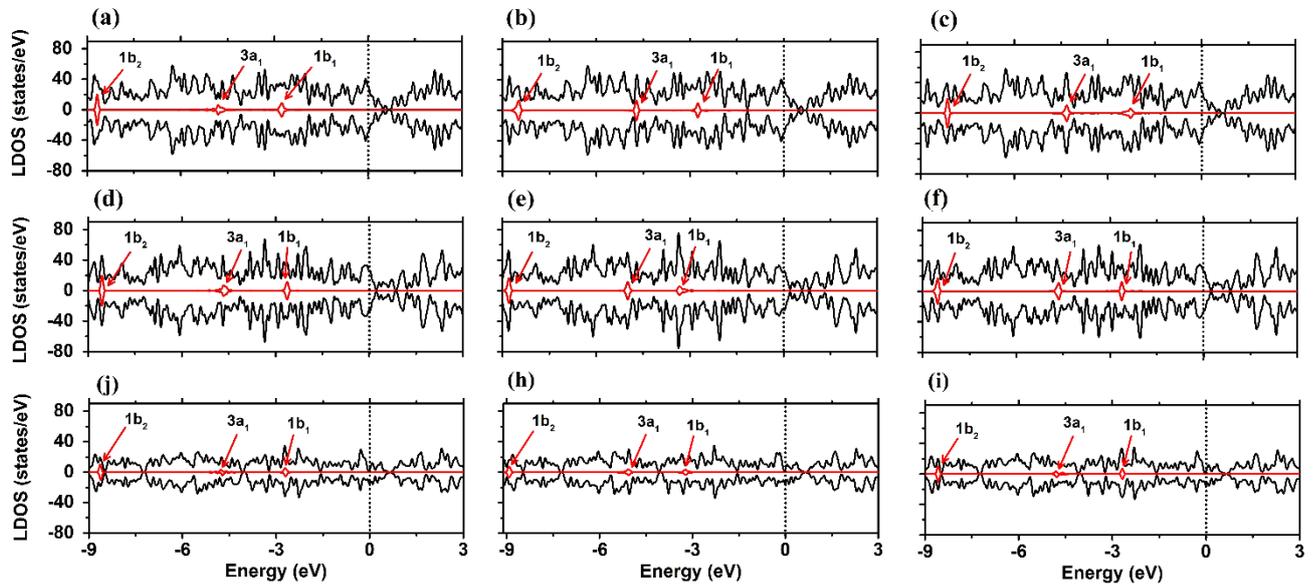

**Figure 3.** LDOS of the $H_2O$ molecule on different regions of the heterostructure. Under 0% strain: (a) graphene region, (e) silicene region, and (c) the graphene/silicene interfacial region. Under -7% strain: (g) graphene region, (d) silicene region, and (f) the graphene/silicene interfacial region. Under 7% strain: (j) graphene, (h) silicene, and (i) the graphene/silicene interfacial region, respectively. The spin-up and -down bands for $H_2O$ are the same and shown by the red line, while the black line represents the total DOS.

### 3.3 Modulation of carrier density and charge transfer

To analyze the electronic interaction of the graphene/silicene heterostructure with the $H_2O$ molecule, the DCD $\Delta\rho(r)$ is calculated. The isosurfaces of the $\Delta\rho(r)$ for the $H_2O$ molecule adsorbed on the different regions of the graphene/silicene heterostructure under 0%, -7% and 7% strains are depicted in figures 4(a)-(c), 5(a)-(c) and 6(a)-(c), respectively.

For the planar surface, there is a depletion of electrons in the $H_2O$ molecule and an accumulation of electrons in the nearest C atoms within the graphene region (figure 4(a)). The $H_2O$ molecule clearly donates electrons to the graphene with around 0.024 *e* per molecule. This result is consistent with the previous DFT study [44], where $H_2O$ was found to be a donor on graphene. However, in a real experiment, the adsorbed water shows the acceptor character [47] due to the time-average behavior of water adsorption with more binding possibilities. The charge-transfer analysis for the $H_2O$ adsorbed on the silicene region of the planar structure (figure 4(b)) shows that the molecule accepts about 0.092 *e*. In the case when the $H_2O$ adsorbed on the planar graphene/silicene interfacial region (figure 4(c)), the total amount of transferred charge is -0.057 *e*. Interestingly, electrons accumulate in the nearest C atoms, while the Si atoms donate electrons to the $H_2O$.

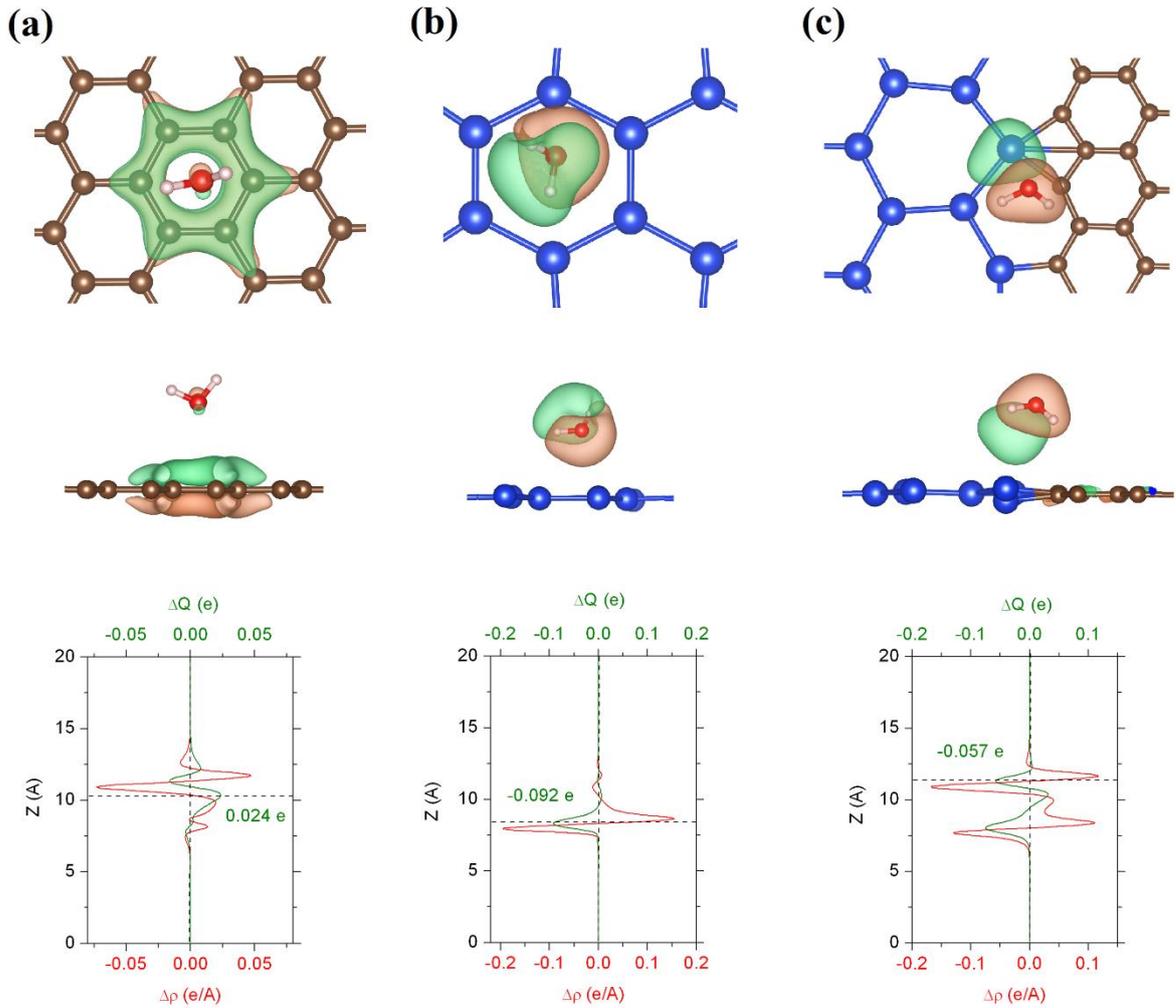

**Figure 4.** Top and side views of the 0.02Å$^{-3}$ DCD isosurface (the green/orange color denotes depletion/accumulation of electrons), the plane-averaged DCD Δρ(z) (red line) and the amount of transferred charge ΔQ(z) (green line) for the H$_2$O molecule adsorbed on (a) graphene, (b) silicene, and (c) graphene/silicene interfacial regions of the planar graphene/silicene sheet.

The compressive strain of -7% leads to an increase in the chemical activity of the H$_2$O molecule, and the total amount of transferred charge from the molecule to the C atoms of the graphene region (figure 5(a)) is 0.035 *e*. Figure 5(b), where the H$_2$O is adsorbed in the silicene region of the compressed structure, clearly indicates an accumulation of electrons in the Si atoms. Thus, H$_2$O serves as a donor in this case where charge transfer from the molecule to the surface is 0.181 *e*. A significant increase of the charge transfer, up to -0.355 *e*, is found for the case of H$_2$O adsorption on the interfacial region of the compressed structure (figure 5(c)), and the main charge transfer is still observed for the C atoms. The obtained results suggest that the carrier density and the charge transfer between the H$_2$O molecule and different regions of the in-plane graphene/silicene heterostructure, as well as a donor/acceptor ability of the H$_2$O, can be significantly tuned by compressive strain.

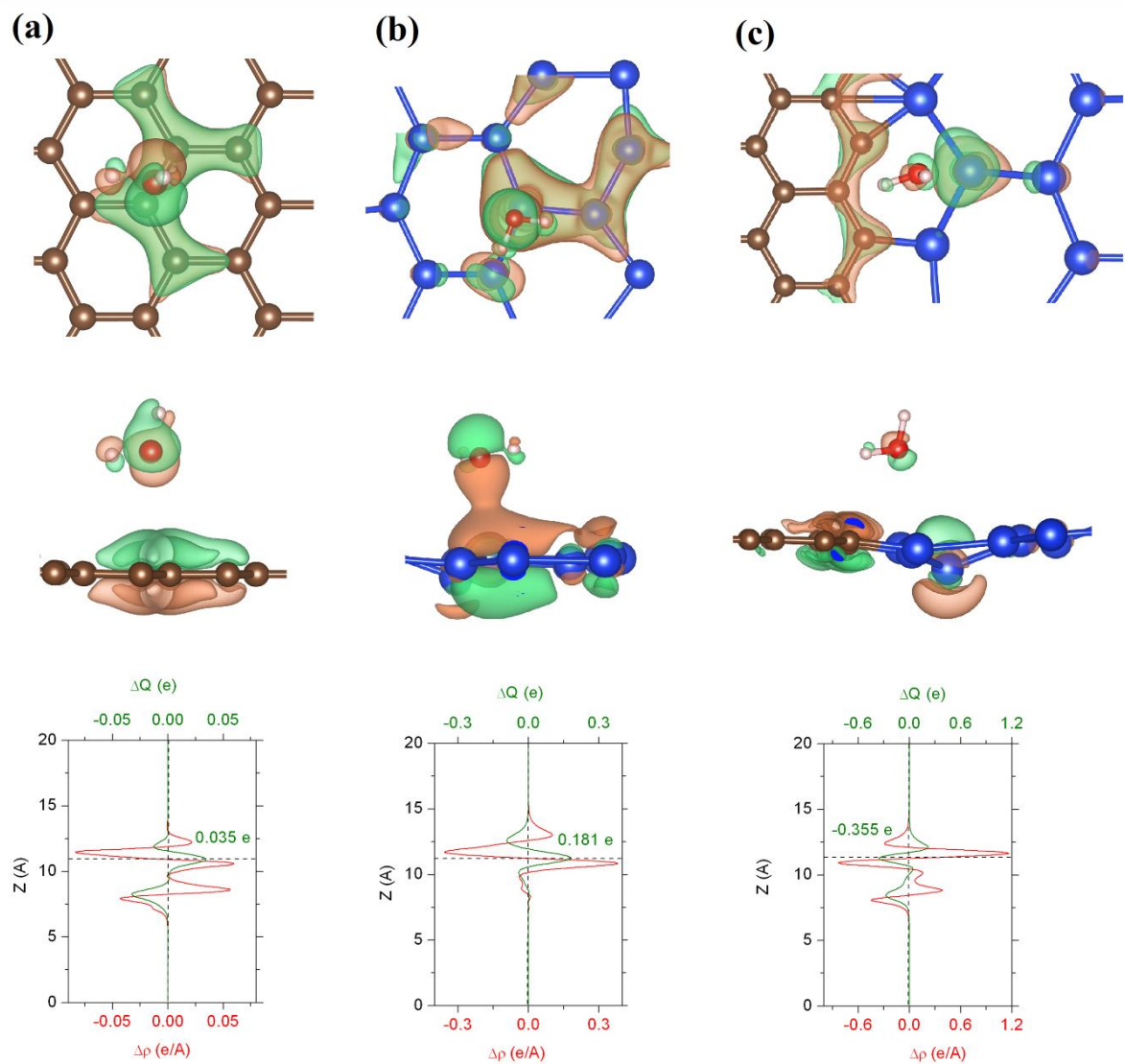

**Figure 5.** Top and side views of the 0.035Å$^{-3}$ DCD isosurface (the green/orange color denotes depletion /accumulation of electrons), the plane-averaged DCD $\Delta\rho(z)$ (red line), and the amount of transferred charge $\Delta Q(z)$ (green line) for the $H_2O$ molecule adsorbed on (a) graphene, (b) silicene, and (c) graphene/silicene interfacial regions of the compressed graphene/silicene sheet.

The tensile strain of 7% also leads to the change in the chemical activity of the $H_2O$ molecule. In particular, the total amount of transferred charge from the C atoms of the graphene region to the molecule (figure 6(a)) is 0.028 e, clearly indicating that $H_2O$ serves as an acceptor. Figure 6(b), where the $H_2O$ is adsorbed on the silicene region of the stretched structure, indicates that molecule donates about 0.160 e to the surface, signifying that $H_2O$ serves as a donor. For the case of $H_2O$ adsorption on the interfacial region of the stretched structure (figure 6(c)) the main charge transfer, up to 0.050 e, is still observed for the $H_2O$ molecule. The obtained results suggest that the carrier density and the charge transfer between the $H_2O$ molecule and different regions of the in-plane graphene/silicene heterostructure, as well as a donor/acceptor ability of the $H_2O$, can be tuned by tensile strain.

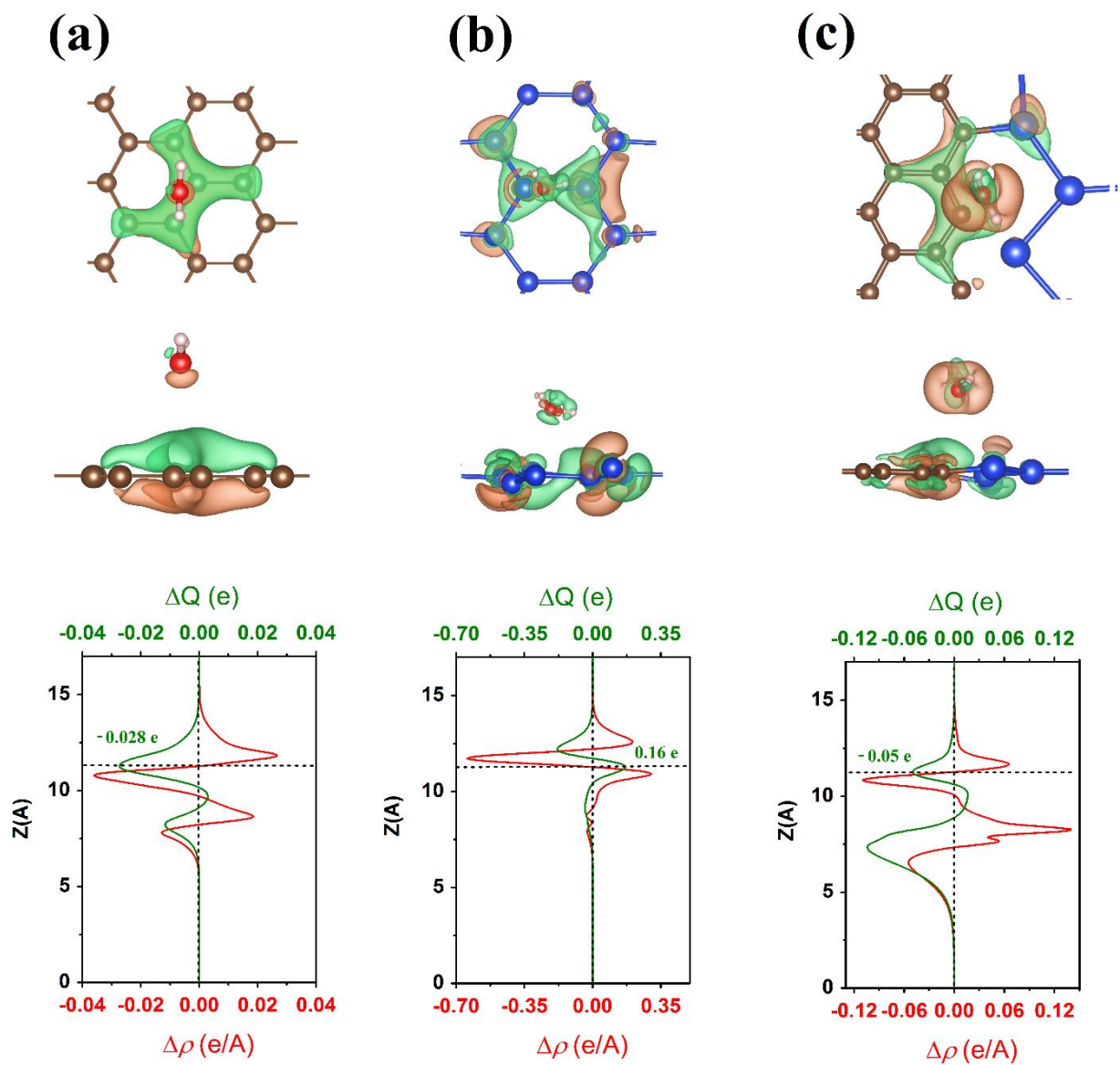

**Figure 6.** Top and side views of the 0.015Å$^{-3}$ DCD isosurface (the green/orange color denotes depletion /accumulation of electrons), the plane-averaged DCD Δρ(z) (red line), and the amount of transferred charge ΔQ(z) (green line) for the H$_2$O molecule adsorbed on (a) graphene, (b) silicene, and (c) graphene/silicene interfacial regions of the stretched graphene/silicene sheet.

**3.4 The effects of the BN-substrate on the chemical activity of the graphene/silicene heterostructure**

To understand the effect of substrate on the chemical activity of the graphene/silicene heterostructure, we choose the h-BN monolayer as a substrate [48, 49]. Specifically, we focus on how the substrate affects the adsorption energies and the charge transfer between the graphene/silicene heterostructure and the adsorbed H$_2$O molecule. For the heterostructure supported by the substrate without applying strain, the most stable configurations for the H$_2$O molecule adsorbed on different regions of the heterostructure are given in figures 7(a), (b) and (c), respectively.

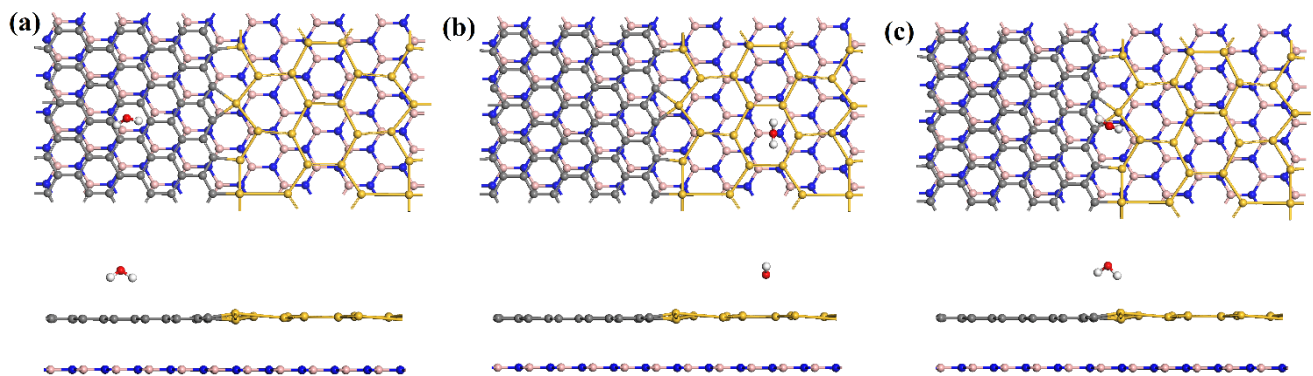

**Figure 7.** The most stable adsorption positions of the H₂O molecule on the graphene/silicene heterostucture supported by the BN layer without applying strain: (a) graphene, (b) silicene, and (c) the graphene/silicene interfacial regions. The balls in gray, yellow, white, red, blue and pink represent carbon, silicon, hydrogen, oxygen, nitrogen and boron atoms, respectively.

In the case of H₂O adsorbed on the graphene region (figure 7(a)), both of the O−H bonds are disposed at the angle of around 45° to the surface and located directly above the ridge of graphene, which is similar to the case in which H₂O is adsorbed on the graphene region of the free-standing surface. The value of $E_a$ is −0.152 eV and the total amount of transferred charge from the molecule to the surface is 0.024 e. Figure 7(b) shows the H₂O molecule adsorbed on the silicene region of the substrate-supported heterostructure, in which the O−H bonds are disposed at the angle of around 45° to the surface and located directly above the ridge of silicene. The value of $E_a$ is −0.140 eV and the total amount of transferred to the molecule is 0.092 e. Figure 7(c) presents the H₂O adsorbed on the graphene/silicene region. It is seen that the molecule is located directly above the C-Si bond of the graphene/silicene site, and both O−H bonds are disposed at the angle of around 45° to the surface. The value of $E_a$ is −0.175 eV and the total amount of transferred to the molecule is 0.057 e.

It is found that the BN-substrate has a small effect on the adsorption energy $E_a$, when H₂O molecule is adsorbed on the graphene region of the graphene/silicene heterostructure. However, for the substrate-supported heterostructure, the $E_a$ is slightly higher for the case of H₂O adsorbed on the silicene region and significantly lower for the case where H₂O is adsorbed on the graphene/silicene interfacial regions, compared with the $E_a$ of the free-standing heterostructure. The charge transfer analysis reveals that the presence of the BN-substrate significantly influences the donor/acceptor ability of H₂O molecule upon its adsorption on the graphene/silicene heterostructure and may cause an increase/decrease of the charge transfer between the H₂O molecule and the heterostructure. Table 2 summarizes the energetics data of the H₂O molecular adsorption on the substrate-supported and free-standing graphene/silicene heterostructure.

Table 2. Adsorption energy ($E_a$) and charge transfer ($\Delta q$) from $H_2O$ molecules to different regions of the planar free-standing and BN-supported graphene/silicene heterostucture. Note that a positive $\Delta q$ indicates the transfer of electrons from the molecules to the graphene/silicene heterostucture.

| Graphene/silicene heterostructure | Heterostructure regions | $E_a$ (eV) | $\Delta q$ (e) | Molecule |
|---|---|---|---|---|
| On the BN-substrate | graphene | -0.152 | 0.024 | donor |
|  | silicene | -0.140 | -0.092 | acceptor |
|  | graphene/silicene | -0.175 | -0.057 | acceptor |
| Free-standing | graphene | -0.146 | -0.014 | acceptor |
|  | silicene | -0.150 | 0.011 | donor |
|  | graphene/silicene | -0.144 | -0.022 | acceptor |

## 4. Conclusions

By using the first-principles calculations the strain effects on the electronic structure of the in-plane graphene/silicene heterostructure are studied. Within the strain range from -7% (compression) to 7% (tension) the considered heterostructure is always metallic with the Dirac point being located above the Fermi level.

The investigation of the strain effects on the chemical activity of the in-plane graphene/silicene heterostructure upon interaction with $H_2O$ molecule reveals that compressive strain is able to promote the adsorption of $H_2O$ molecule and increase the charge transfer, signifying an enhanced chemical activity. Furthermore, compressive and tensile strains are found to be able to modulate the charge transfer between the $H_2O$ molecule and graphene/silicene surface, potentially allowing the control of polarity and concentration of charge carriers. In addition, the effect of the BN-substrate on the chemical activity of the in-plane graphene/silicene heterostructure upon interaction with the $H_2O$ molecule is considered. It is found that the BN-substrate significantly influences the donor/acceptor ability of $H_2O$ molecule upon its adsorption on the graphene/silicene heterostructure and may cause an increase/decrease of the charge transfer between the $H_2O$ molecule and heterostructure. This study provides deep insights into the effects of strain and water molecule on the electronic properties and chemical activity of the in-plane graphene/silicene heterostructure, which may be useful for its potential device applications.

# ACKNOWLEDGMENTS

The authors gratefully acknowledge the financial support from the Ministry of Education, Singapore (Academic Research Fund TIER 1 − RG128/14), the Agency for Science, Technology and Research (A*STAR), Singapore, and the use of computing resources at the A*STAR Computational Resource Centre, Singapore. This work was supported in part by a grant from the Science and Engineering Research Council (152-70-00017). Sergey V. Dmitriev acknowledges financial support from the Russian Science Foundation grant N 14−13−00982.